\documentclass{aa} 
\usepackage[utf8]{inputenc} 
\usepackage{threeparttable}
\usepackage[normalem]{ulem}
\usepackage{txfonts}
\usepackage{hyperref}
\usepackage{amsmath}
\usepackage{xcolor}

\newcommand{\DSFMS}{\Delta_\mathit{SFMS}}
\newcommand{\SFR}{\mathrm{SFR}}

\newcommand{\Gyr}{\mathrm{Gyr}}
\newcommand{\Myr}{\mathrm{Myr}}

\newcommand{\dex}{\mathrm{dex}}
\newcommand{\Msun}{\mathrm{M}_\odot}

\newcommand{\kmsMpc}{\mathrm{km}\,\mathrm{s}^{-1}\,\mathrm{Mpc}^{-1}}

\newcommand{\mgfep}{$\rm [MgFe]^\prime$}
\newcommand{\mgtwofe}{$\rm [Mg_2Fe]$}
\newcommand{\SFRDR}{$\rm SFR^{DR8}\,$}
\newcommand{\SFRS}{$\rm SFR^{S16}\,$}

\defcitealias{Mattolini2025}{M25}

\begin{document} 
\title{A stellar Fundamental Metallicity Relation for galaxies}
    \subtitle{The dependence of stellar metallicity on stellar mass and star formation rate in the Local Universe}

   \author{D. Mattolini,
          \inst{1,2}
          S. Zibetti,
          \inst{2}
          A. R. Gallazzi,
          \inst{2}
          L. Scholz-Díaz
          \inst{2}
          }
\authorrunning{Mattolini et al.}
   \institute{Dipartimento di Fisica ,
            Universit\`a di Trento, Via Sommarive 14, I-38123 Povo (TN), Italy
        \and
            INAF-Osservatorio Astrofisico di Arcetri, Largo Enrico Fermi 5, I-50125 Firenze, Italy
             }

\date{Received $27$ July 2026 / Accepted $15$ November 2026}
\abstract
{We investigate the connection between current star formation activity and mean stellar metallicity as a function of stellar mass, using integrated properties of 74\,350 galaxies at $0.005\leq z\leq0.22$ with spectral S/N $\geq20$ from the Sloan Digital Sky Survey Data Release 7. Using aperture-corrected stellar masses and light-weighted mean stellar metallicities ($Z_\star$) we find that $Z_\star$ depends primarily on stellar mass, following a mass--metallicity relation (MZR) whose slope flattens with increasing mass. We also find a highly significant secondary anticorrelation between $Z_\star$ and star formation rate (SFR) at fixed mass. By deriving the MZR in bins of distance from the star-forming main sequence, we characterise a stellar fundamental metallicity relation (sFMR) in which more actively star-forming galaxies i) have systematically lower $Z_\star$ at fixed mass, and ii) follow progressively steeper MZRs, which converge at the high-mass end.

The sFMR is qualitatively similar in shape to the well-known gas-phase FMR, but differs in normalisation and, notably, in its high-mass behaviour: in each SFR bin, the gas-phase MZR flattens at high masses, whereas the stellar MZR keeps rising. A consistently derived sFMR at $z\sim0.7$ shows a similar shape, albeit with lower normalisation, indicating that the coupling between star formation and chemical enrichment was already in place at earlier epochs. Taken together, these results suggest that the chemical evolution of gas and stars in galaxies is smoothly coupled to their star formation history over timescales of several Gyr, disfavouring a dominant role of stochastic star formation episodes.}

\maketitle
\section{Introduction}\label{sec:Introduction}
In the Local Universe, galaxies exhibit strong correlations between stellar mass and both their stellar and gas-phase metallicities, giving rise to the gas and stellar metallicity-mass relations \citep[MZR; e.g.][see also the review by \citealp{Maiolino2019}]{Tremonti2004, Gallazzi2005, Gallazzi2008, Ellison2008, Panter2008, Mannucci2010, Pasquali2010, Zahid2013, Zahid2017, Peng2015, Sanchez2018, Curti2020, Trussler2020, Looser2024}.
These relations are shaped by the interplay between metal production, metal-poor gas inflows and metal-rich stellar/AGN driven gas outflows \citep[][]{DiMatteo2005, Tumlinson2011, Cicone2014, Chisholm2018}, that govern the baryon cycle and galaxies assembly histories.
They represent fundamental observational constraints to test predictions from theoretical models of galaxy formation and evolution, such as cosmological hydrodynamical simulations \citep[e.g.][]{Vogelsberger2014, Schaye2015, Dave2016, Springel2018} and semi-analytic models (e.g. \citealp{Somerville2008, Guo2011, DeLucia2014}; see also the reviews by \citealp{Somerville_Dave2015, Vogelsberger2020}).

The information provided by the gas- and stellar-MZRs are complementary.
The average stellar metallicity reflects the full chemical enrichment history of a galaxy, as imprinted in the fossil record stellar component.
Conversely, the gas-phase metallicity traces the current composition of the ionised interstellar medium (ISM) that is being inherited by the present generation of stars.
Therefore, variations of the gas- and stellar-MZRs are sensitive to processes happening on different timescales.

In the last two decades, studies of both local and high redshift galaxies have shown that the gas-phase MZR has a secondary dependence on galaxies' star-formation activity, in particular the current star formation rate (SFR).
Such dependence results in a $M_\star-Z_{\rm gas}-\rm SFR$ relation according to which, at any given mass, more actively star-forming galaxies exhibit on average a lower gas-phase metallicity compared to their less star-forming counterparts \citep[][]{Ellison2008, Lara-Lopez2010, Mannucci2010, Hunt2012, Salim2014, Sanders2018, Cresci2019, Curti2020}.
Due to the lack of redshift evolution up to $z\sim2.5$ this relation is usually referred to as the Fundamental Metallicity Relation \citep[FMR; see ][]{Mannucci2010}.

A possible interpretation of the gas-phase FMR is offered by self-regulation (or "gas-regulator") models \citep[e.g.][]{Dave2011, Dayal2013, Lilly2013, DeRossi2015}, in which galaxy chemical evolution is governed by a balance between metal-poor gas accretion, metal-enriched outflows, and in situ metal production via stellar nucleosynthesis. 
Inflow of pristine gas simultaneously dilutes the metal content of the ISM and fuels star formation (and, in massive galaxies, black hole accretion), which increases metal production through stellar nucleosynthesis. 
The resulting energetic feedback — from supernovae and, at the high-mass end, from the AGN — drives metal-enriched outflows that remove metals from the ISM and suppress further star formation, closing the regulatory loop.
The FMR thus represents a powerful constraint to test star formation, feedback, and metal enrichment processes in theoretical models of galaxy evolution.

More recently, the analysis of the secondary dependence of metallicity on galaxies' star formation activity has been explored also for the stellar component.
At low-redshift it has been found that star-forming and passive galaxies follow distinct MZRs \citep[e.g.][]{Peng2015, Trussler2020, Gallazzi2021, Mattolini2025} and, more recently, \cite{Looser2024} found a secondary continuous dependence of stellar metallicity on the SFR for a sample of $7\,323$ MaNGA galaxies, defining a stellar-FMR analogous to the one observed for the gas-phase metallicity.
For simplicity and consistency with earlier studies in the literature \citep[e.g.,][]{Fontanot2021, Garcia2024, Looser2024}, and in analogy with the gas-phase relation, we refer to this as the stellar-FMR (sFMR).
We note, however, that this relation may vary with redshift, as discussed in Sect. \ref{sec:Discussion} \citep[see also][]{Fontanot2021}.
At higher redshift, \cite{Gallazzi2025PaperII} reported a similar $M_\star-Z_{\star}-\rm SFR$ relation at $z\sim0.7$, based on a sample of $552$ galaxies from the LEGA-C survey.
Although quantitatively different from the gas-FMR (gFMR), the stellar-FMR found both locally and at higher redshift display qualitatively similar trends: more actively star-forming galaxies have lower stellar metallicities compared to less active galaxies of the same mass.

The presence of a sFMR has been primarily interpreted as evidence for galaxy starvation \citep[see e.g.][]{Larson1980, Balogh2000, vandenBosch2008}, which is characterised by a reduction/stop of metal-poor gas inflow from the inter galactic medium.
In this scenario a progressively suppressed star-formation activity happens in a progressively enriched environment, yielding an anticorrelation between SFR and stellar metallicity \citep[see][]{Peng2015, Trussler2020, Looser2024}. Stellar or AGN driven galactic outflows have also been proposed as equally important processes in shaping the history of metal production, coupled with starvation \citep{Trussler2020} or with inefficient conversion of gas into stars in currently star forming galaxies \citep{Spitoni2017}.
The influence of these processes is likely dependent on the galaxy potential well, as traced by galaxy size \citep{Barone2018, Barone2020, Vaughan2022} and stellar surface mass density \citep[][]{Zibetti2022}.
Metal cycle regulation processes coupled with structural differences between passive and star-forming galaxies could give rise to the observed difference in stellar metallicity with SFR.

In this work, we leverage the newly derived stellar metallicities of \cite{Mattolini2025} (\citetalias{Mattolini2025} hereafter), obtained with state of the art stellar population synthesis (SPS) models and implementing corrections for the Sloan Digital Sky Survey \citep[SDSS][]{York2000} aperture-bias \citep[see][]{Zibetti2026}, to quantify the dependence of stellar metallicity on both stellar mass and star formation activity.
Our study expands the analysis of the sFMR in the local Universe to a sample of $74~350$ galaxies, a factor 10 larger with respect to the most comprehensive work conducted so far by \cite{Looser2024}.
Notably, the use of a consistent methodology with the one adopted by \cite{Gallazzi2025PaperII} in the LEGA-C sample at $z\sim0.7$ enables the study of the evolution of the sFMR over cosmic times.

The paper is organised as follows.
In Sect. \ref{sec:Data_analysis} we present the data and methodology used in our study.
In Sect. \ref{sec:Results} we show our results, and discuss their implications in Sect. \ref{sec:Discussion}.
A summary and the conclusions of our work are presented in Sect. \ref{sec:Conclusions}.
Throughout the paper, we assume a $\Lambda\text{CDM}$ cosmology with $H_0=70\,\kmsMpc$, $\Omega_\mathrm{m}=0.3$, and $\Omega_\Lambda=0.7$.

\section{Data analysis}\label{sec:Data_analysis}

In this work we study the mutual dependence of stellar mass, mean stellar metallicity and SFR for a sample of $89\,852$ galaxies at $0.005\leq z\leq0.22$ drawn from the Data Release 7 \citep{Abazajian2009} of the SDSS \citep[][]{York2000}.
Stellar masses and stellar metallicities are drawn from recently published catalogues of stellar population parameters by \citetalias{Mattolini2025}.
Since light-weighted properties are less dependent on the star formation history (SFH) and chemical enrichment history (CEH) assumptions \citep[e.g.][]{Serra2007, Zibetti2017}, in this work we use the light-weighted estimates of stellar metallicities, as defined in equation 3 of \citetalias{Mattolini2025}\footnote{Mass-weighted metallicities lie systematically below light-weighted ones, typically by a few 0.01 dex and by up to $\sim0.15~\dex$ in the most actively star-forming galaxies. This is expected, since light-weighted estimates are biased towards younger, more metal-rich stellar populations. These differences are small and do not affect the trends found in this work.}

These parameters were derived within the Bayesian framework BaStA\footnote{See also \url{https://www.basta.inaf.it/}} \citep{Gallazzi2005, Zibetti2017}, in which the prior is defined by a pre-computed library of $500\,000$ composite stellar populations (CSPs).
For each model we computed the likelihood of the data, $\mathcal{L}\propto\exp(-\chi^2/2)$.
The posterior probability distribution function (PDF) of the stellar metallicity was then obtained by weighting each model by its likelihood and marginalising over all other model parameters.
We adopt the median of the posterior PDF as the fiducial value and half the difference between its 16th and 84th percentiles as the uncertainty.

The CSP models were built from the latest (2019) version of the \citet{Bruzual2003} simple stellar populations (SSPs; CB19 hereafter), based on PARSEC evolutionary tracks \citep{Bressan2012, Marigo2013} and MILES stellar spectra \citep{Sancez-Blazquez2006}.
These SSPs follow the Milky Way $[\alpha/\mathrm{Fe}]$ abundance pattern and are provided at 221 time steps between 0 and $14~\Gyr$, for a grid of 17 metallicities between $[\mathrm{M/H}]=0$ and $[\mathrm{M/H}]=0.60$ \citep[see appendix~A of][for details]{Sanchez2022}.
The CB19 SSPs were combined with delayed Gaussian star formation histories (SFHs; \citealt{Gavazzi2002}),
\begin{equation}
\mathrm{SFR}_\tau(t)\propto\frac{t-t_\mathrm{form}}{\tau}\exp\left[-\frac{(t-t_\mathrm{form})^2}{2\tau^2}\right],
\end{equation}
on top of which random bursts are superimposed to account for stochastic star formation.
Following \citet{Zibetti2017}, the chemical enrichment histories (CEHs) follow a mass-dependent, monotonically rising parametric form,
\begin{equation}
Z_\star(t)=Z_{\star,\mathrm{final}}-\left(Z_{\star,\mathrm{final}}-Z_{\star,0}\right)\left[1-\frac{M(t)}{M_\mathrm{final}}\right]^\alpha,
\end{equation}
where $M(t)$ is the stellar mass formed up to time $t$, $M_\mathrm{final}$ is the total stellar mass formed over the SFH, $Z_{\star,0}$ and $Z_{\star,\mathrm{final}}$ are the initial and final metallicities, and $\alpha$ sets how fast chemical enrichment proceeds relative to star formation.
Dust attenuation is modelled with the two-component model of \citet{Charlot2000}.

Stellar masses and light-weighted stellar metallicities were inferred by comparing model predictions with absorption-index measurements and SDSS $ugriz$ Petrosian photometry.
To mitigate the age--metallicity degeneracy, we used indices that are primarily sensitive to either age or metallicity.
As age-sensitive features we used the H$\beta$, H$\delta$, and H$\gamma$ absorption lines and the 4000\,\AA\ break.
As metallicity-sensitive indices we used the composite indices \mgfep{} and \mgtwofe{} \citep{Thomas2003, Bruzual2003}, which are robust metallicity indicators with minimal dependence on $[\alpha/\mathrm{Fe}]$.

SDSS spectra were obtained through fixed-size fibres, which cover different fractions of galaxies of different apparent size. Combined with radial gradients in stellar population properties, this introduces the so-called aperture bias \citep{Kauffmann2003, Gallazzi2005, Peng2015, Trussler2020}.
In \citetalias{Mattolini2025}, the absorption indices were aperture-corrected as a function of the fibre index strength, $(g-r)$ colour, Petrosian half-light radius, and $r$-band absolute, following \cite{Zibetti2026}.
Because the corrections account for the more metal-poor outer regions not covered by the fibre, they lower $Z_\star$ in a mass-dependent way, by up to $\sim0.3~\dex$ at the low-mass end.
For a comprehensive analysis of the systematics associated with the choice of SPS models and with the aperture corrections, we refer the reader to \citetalias{Mattolini2025}.

The SFR estimates used in the following analysis are drawn from the GALEX–SDSS–WISE Legacy Catalog (GSWLC)  \citep{Salim2016}, and are obtained via UV/optical SED fitting.
We matched the \citetalias{Mattolini2025} and \cite{Salim2016} catalogues by sky coordinates with a tolerance of $1''$.
Filtering for average spectral $\rm S/N\geq20$ per pixel and available estimates of SFR, we obtain a final sample of $74\,350$ galaxies.
As an alternative option to validate our results and assess possible systematic effects, we also considered $\rm H\alpha$-based SFR estimates of SDSS DR8 galaxies \citep[][]{Aihara2011} provided building upon \cite{Brinchmann2004} and \cite{Salim2007}.
We discuss the impact of such a different choice of SFR estimates in Appendix \ref{app:sfr_comparison}.
Throughout the rest of the paper we will focus on the SFR by \cite{Salim2016}.

In order to quantify the secondary dependence of stellar metallicity on SFR, we bin our sample based on the distance ($\Delta_\mathit{SFMS}$) from the star forming main sequence (SFMS) in the SFR vs $M_\star$ plane \citep[following a popular approach in the literature, e.g.][\citetalias{Mattolini2025}]{Trussler2020, Gallazzi2021, Gallazzi2025PaperII, Looser2024}.
To ensure full consistency, we defined the SFMS based on our data as a linear fit to the locus of star-forming peaks in the $\log(M_\star)-\log(\mathrm{SFR})$ plane, derived from the SFR distributions in $0.1~\mathrm{dex}$ wide stellar mass bins.
We define seven bins of: 
\begin{equation}
    \Delta_\mathit{SFMS}\equiv\log(\textrm{SFR}~[\Msun \textrm{yr}^{-1}])-\log(\textrm{SFR}_\textrm{MS}(M_\star))~,
\end{equation}
where
\begin{equation}\label{eq:SFMS_fit}
    \log(\textrm{SFR}_\textrm{MS}(M_\star))=0.61\log(M_\star/\Msun)-6.11~
\end{equation}
is our fit of the SFMS.
Additionally to this $\DSFMS$-based selection, we analysed the MZR dividing galaxies in bins of SFR and sSFR, enabling a more comprehensive comparison with the literature.
\begin{figure}
\centering
\includegraphics[width=\hsize]{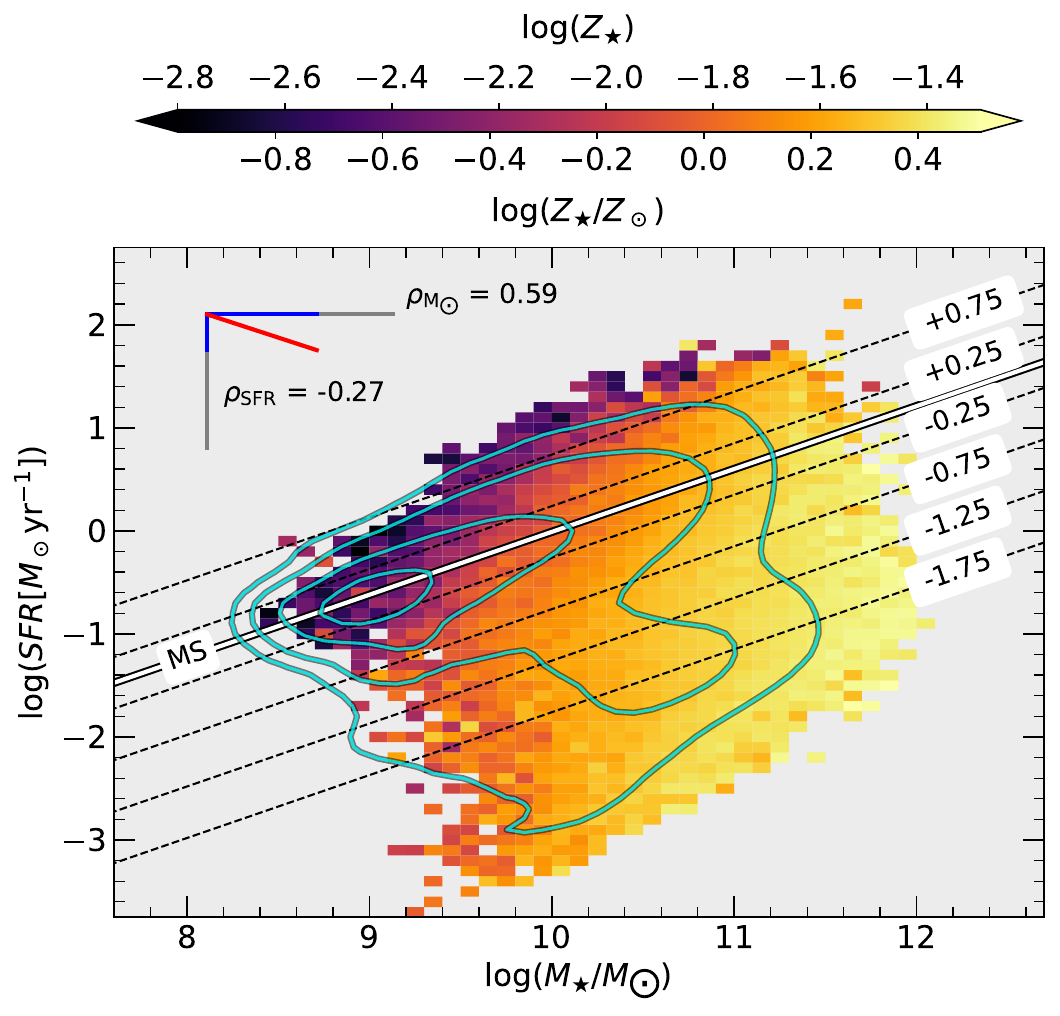}
  \caption{$M_\star-\SFR$ distribution colour-coded by the median $\log(Z_\star)$.
  The cyan density levels enclose $16\%$, $50\%$, $84\%$, and $97.5\%$ of the complete and volume-corrected galaxy sample, obtained by smoothing the corresponding distribution using a Gaussian KDE.
  The oblique solid white line identifies our fit of the SFMS, while the oblique dashed lines separate different $0.5~\dex$-wide bins of $\DSFMS$.
  The blue lines in the top left corner identify the partial correlation coefficients, and the red line identifies the direction of maximal increase of $\log(Z_\star)$ in the plane.}
    \label{fig:Mstar-SFR_cc_Zstar}
\end{figure}

\section{Results}\label{sec:Results}
\begin{figure*}[!ht]
\centering
\includegraphics[width=.8\hsize]{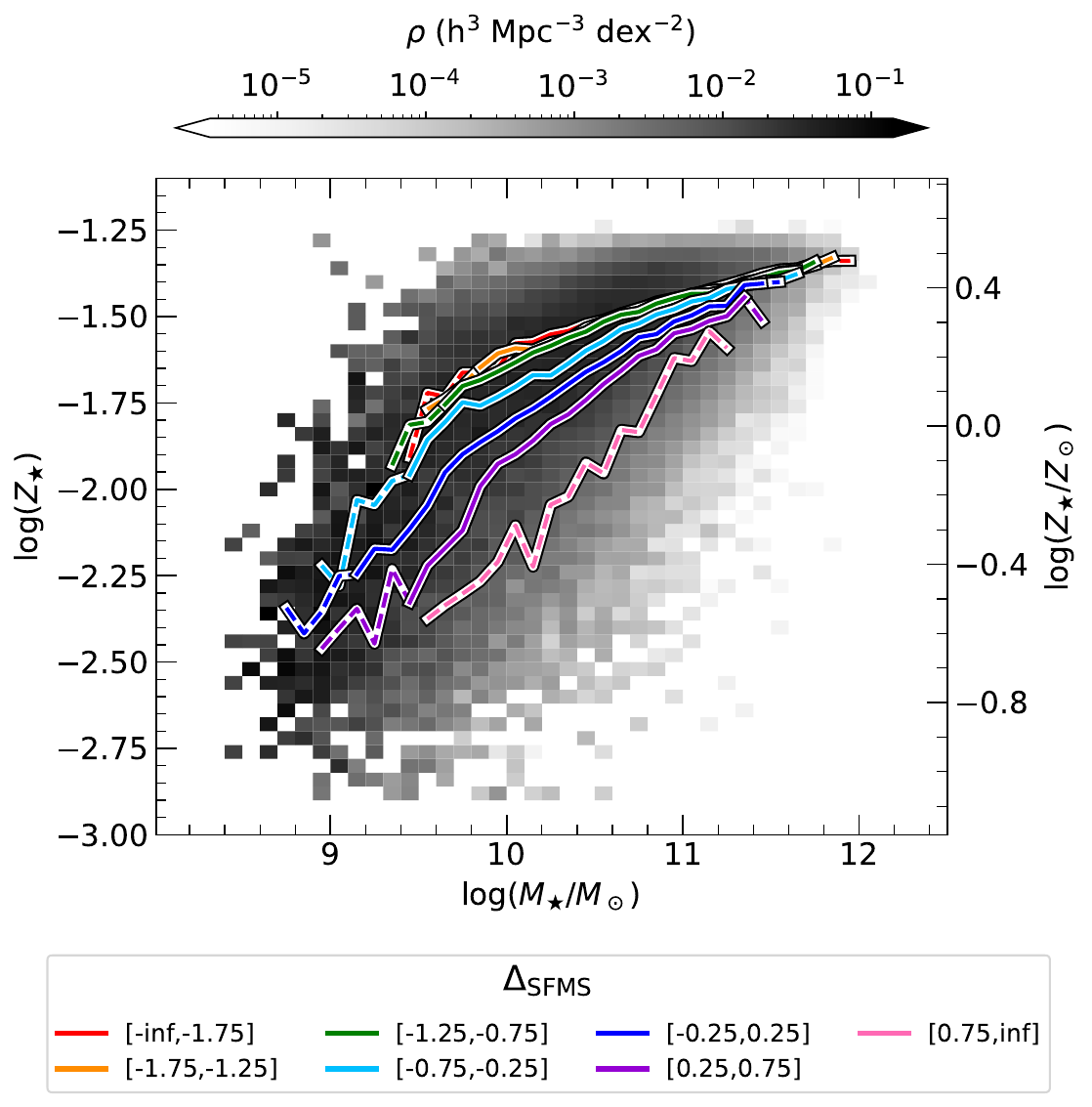}
  \caption{The local Universe stellar-FMR.
  The coloured lines identify the median MZRs of galaxies in different bins of $\DSFMS$.
  The solid (dashed) lines identify those bins with at least 50 (10) galaxies per bin.
  The grey 2D histogram represents the number density of the complete galaxy sample in the $M_\star-Z_\star$ plane.
  The dependence of $Z_\star$ on galaxies star formation activity result in MZRs with larger normalisation and shallower slopes with decreasing star formation activity, similar to the gas-phase FMR.}\label{fig:FMR_dist_sfr}
\end{figure*}
We start by analysing how the average stellar metallicity changes as a function of $M_\star$ and $\mathrm{SFR}$. In Fig. \ref{fig:Mstar-SFR_cc_Zstar}, each 2D-bin in the $M_\star-\mathrm{SFR}$ plane is colour-coded with the median value of $\log(Z_\star)$ of the galaxies in that bin.
The cyan contours identify the isodensity levels of the volume-corrected number density distribution of the galaxies, while the solid white line identifies our fit of the SFMS, as defined in eq. \eqref{eq:SFMS_fit}.
We visually identify a gradient for stellar metallicity that is not parallel to either of the axes, indicating that $Z_\star$ depends on both $M_\star$ and $\SFR$.
Following \cite{Scholz-Diaz2022, Scholz_Diaz2024}, in order to quantify the gradient while controlling for the intrinsic correlation of $M_\star$ and $\mathrm{SFR}$, we compute partial correlation coefficients (based on Spearman rank). Their amplitudes, and the resulting combined vector indicating the gradient of $\log(Z_\star)$ in the plane, are displayed in the top-left corner of Fig. \ref{fig:Mstar-SFR_cc_Zstar} with blue and red lines, respectively.
The partial correlation analysis shows that stellar metallicity is most strongly linked to the total stellar mass of the galaxy, with a coefficient $\rho_{M_\star}=0.59\pm0.01$, reflecting the well-known MZR. 
Nonetheless, the partial correlation coefficient of $Z_\star$ with $\SFR$ $\rho_\SFR=-0.27\pm0.01$ implies a significant secondary dependence of $Z_\star$ on $\SFR$ at fixed $M_\star$.
Both correlations are highly statistically significant (showing p-values below $0.001$), with the quoted uncertainties representing the 95\% confidence intervals.

\begin{figure*}
\centering
\includegraphics[width=\hsize]{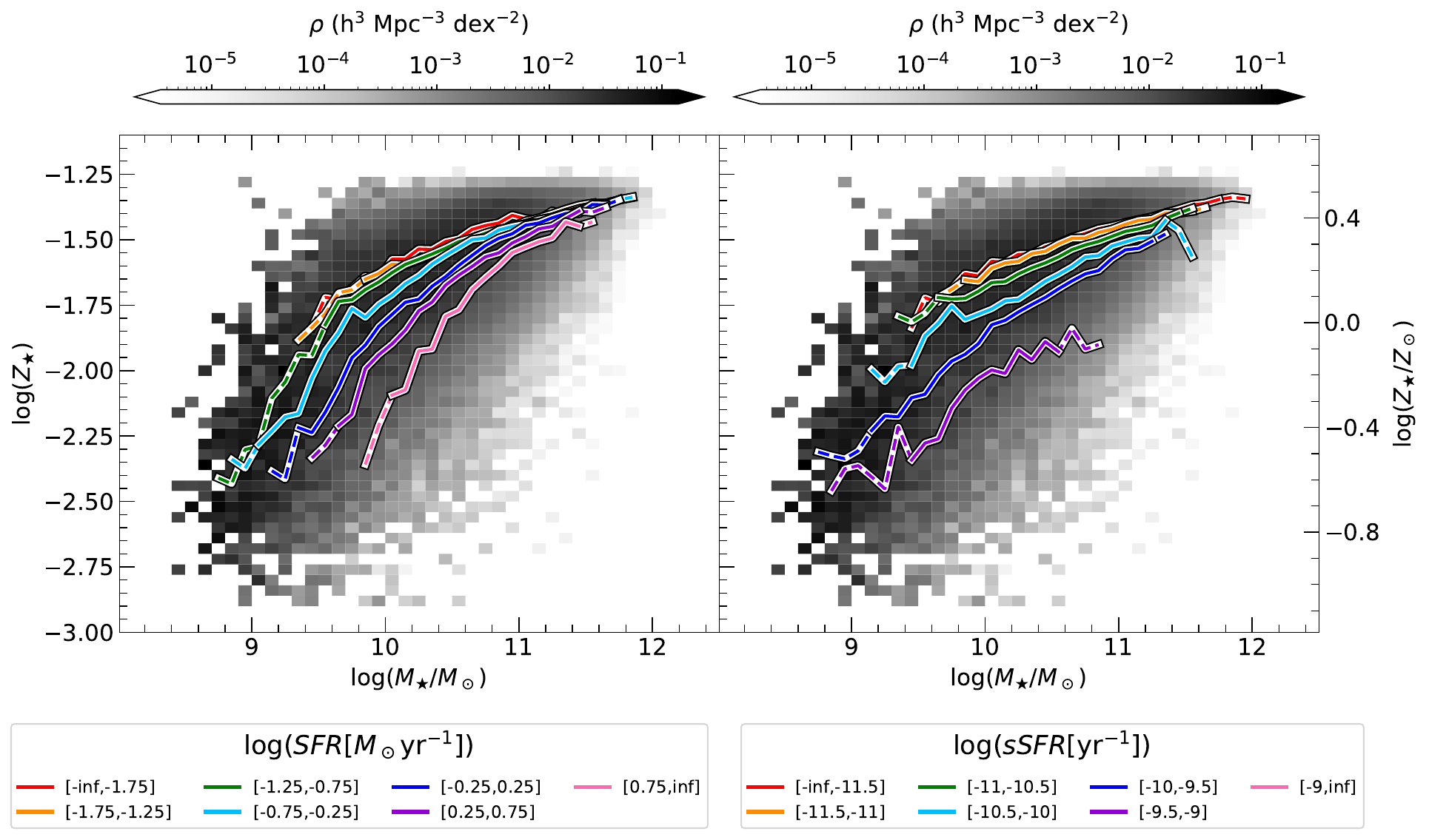}
  \caption{The coloured lines identify the median MZRs of galaxies in different bins of SFR (left panel) and sSFR (right panel).
  The solid (dashed) lines identify those bins with at least 50 (10) galaxies per bin.
  The grey 2D histogram represents the number density of the complete galaxy sample in the $M_\star-Z_\star$ plane.}\label{fig:FMR_sfr_ssfr}
\end{figure*}

The secondary dependence of $Z_\star$ on SFR can be alternatively analysed by deriving the MZR for galaxies in different bins of $\DSFMS$. The dashed diagonal lines in Fig. \ref{fig:Mstar-SFR_cc_Zstar} illustrate the binning in $\DSFMS$ we adopt for this analysis.
In Fig. \ref{fig:FMR_dist_sfr} the median\footnote{All median estimates consider statistical weights to account for volume corrections and incompleteness, as derived in \citetalias{Mattolini2025}.} MZR in each bin of $\DSFMS$ is represented by the coloured lines (as labelled in the legend), where solid and dashed lines are used to indicate mass bins with at least $50$ and between $10$ and $49$ galaxies, respectively.
As a reference, the number density distribution of the galaxy sample in the $M_\star-Z_\star$ plane is represented by the underlying grey 2D histogram.
The most actively star-forming galaxies follow steep MZRs, and have the lowest $Z_\star$ at any fixed $M_\star$.
With decreasing star formation activity galaxies follow progressively flatter MZRs and have larger stellar metallicities at fixed $M_\star$.
The dependence of the MZR upon $\DSFMS$ weakens at lower star-formation activity, becoming negligible for galaxies that deviate below the SFMS by more than $\sim1.25~\dex$.
Moreover, the largest variations of $Z_\star$ at fixed stellar mass across different bins of $\DSFMS$ are found in the low-mass end.
With increasing $M_\star$ the dependence of $Z_\star$ on $\DSFMS$ progressively diminish and becomes negligible above $\sim10^{11}~\Msun$.

Consistent results are found when using $H\alpha$-based SFR estimates (see Appendix \ref{app:sfr_comparison}), and when binning galaxies according to SFR and specific-SFR (sSFR) rather than $\DSFMS$.
As shown in the left panel of Fig. \ref{fig:FMR_sfr_ssfr}, considering bins in SFR the trends of the MZRs are qualitatively the same as for the binning in $\DSFMS$.
When binning is made according to sSFR, we find MZRs with similar slope but decreasing normalisation with increasing sSFR, as shown in the right panel of Fig. \ref{fig:FMR_sfr_ssfr}.
Despite these differences, the observed trend of decreasing $Z_\star$ with increasing star-formation activity at fixed mass is maintained.

To determine the significance of the observed trends, we compare the offsets in metallicity for the MZRs in different bins of star formation activity with the uncertainties on these median relations\footnote{Note that uncertainties on $M_\star$ are negligible with respect to the uncertainties on $Z_\star$ when considering the MZRs.}. 
As a function of $M_\star$, we consider the uncertainty on the MZR $\sigma_Z(M_\star)$ in a given bin of star formation activity as the dispersion about the median (i.e. half of the difference between the 84th and 16th percentile of the $Z_\star$ distribution), divided by the square root of the number of galaxies in each mass bin.
Thanks to the large sample available to our study, despite the relatively large dispersions of $0.1-0.2~\dex$ (driven by the errors in the individual estimates of $Z_\star$), the typical values of $\sigma_Z$ are of the order of $\sim0.01~\dex$, with the largest values of $\sim0.04~\dex$ occurring in the least populated bins. 
We measure separations between the MZRs significantly larger than $\sigma_Z$ (on average a factor $\sim7$), over most of the mass range, except for the three least actively star-forming bins where the MZRs overlap. We can thus conclude that the trends of the MZRs with star formation activity are statistically robust. In Appendix \ref{app:MZRs_bins_SFR} we analyse the dispersions in connection with the typical $Z_\star$ estimate uncertainties in more detail. We also present a test showing that the sFMR is stable (although more noisy) against a more conservative sample selection requiring $\rm S/N>30$.

\section{Discussion}\label{sec:Discussion}
Our analysis demonstrates that in the local Universe, the secondary dependence of $Z_\star$ on SFR yields distinct MZRs when galaxies are categorised by their star formation activity.
Together, these define a manifold in the $M_\star$--$Z_\star$--$\mathit{SFR}$ space, analogous to the gas-phase FMR, which we term the stellar-FMR, following earlier works in the literature \citep[e.g.][]{Looser2024}.
The UV-based SFRs adopted in our analysis trace star formation averaged over the past $\sim 100$ Myr.
In contrast, $Z_\star$ reflects the cumulative chemical enrichment history of the galaxy, retaining the imprint of earlier stellar generations formed over much longer timescales.
Therefore, the observed dependence of $Z_\star$ on the current SFR links processes operating on vastly different timescales.
This connection suggests that the mechanisms shaping a galaxy's chemical evolution and regulating its star formation follow smooth and coherent evolutionary paths.
The existence of a sFMR implies that the current star formation activity of a galaxy is tightly linked to its past star-formation and chemical enrichment history, thus disfavouring strong stochasticity in galaxies' star formation activity.

Notably, galaxies in the three least actively star-forming bins exhibit negligible differences in their MZRs.
Therefore, for galaxies below the green valley, minor variations in SFR are no longer associated with appreciable differences in $Z_\star$, indicating that minor star formation episodes may occur independently of the past star-formation and chemical enrichment history.
Nevertheless, as discussed in details in Appendix \ref{app:sfr_comparison}, the shape and normalisation of the MZR in bins with such small star formation activity may depend on the methodology used to estimate SFR.

Our results are in agreement with past studies in the local Universe which, by dividing galaxies between passive and star-forming, found distinct MZRs (e.g. \citealp{Peng2015, Trussler2020, Gallazzi2021}; \citetalias{Mattolini2025}), with passive galaxies following flatter MZRs with higher $Z_\star$ at a given $M_\star$.
A proposed explanation for such MZRs invokes prolonged periods (some Gyr) of starvation \citep[i.e. a stop/reduction of gas infall from the inter-galactic medium; see e.g.,][]{Larson1980, Balogh2000, vandenBosch2008}.
Possibly coupled with AGN-driven outflows \citep[][]{DiMatteo2005, Schawinski2007, Cicone2014}, starvation can simultaneously shape both the chemical enrichment history and the SFH, thereby coupling them on long timescales \citep[][]{Peng2015, Trussler2020, Looser2024}. 
Alternative explanations attribute these distinct MZRs to the interplay between varying gas accretion timescales and galactic outflow efficiencies, which jointly regulate the replenishment and depletion of the galactic gas reservoir \citep[see][]{Spitoni2017}.

Recent studies have extended this analysis to finer bins of star formation activity \citep[e.g.][]{Zahid2017, Looser2024}.
Although MZR slopes and normalisations depend on the method used to estimate $Z_\star$ (see, e.g., Sect.~5 of \citetalias{Mattolini2025} for the biases introduced by different SPS models), the relative trends of $Z_\star$ with SFR reported in these studies are qualitatively consistent with ours.
The \citet{Looser2024} MZRs are, however, offset by several tenths of a dex towards lower metallicities with respect to most literature determinations (e.g. \citealp{Gallazzi2005, Gallazzi2021, Panter2008, Gonzalez-Delgado2014, Trussler2020, Neumann2021}; \citetalias{Mattolini2025}).
Neither aperture effects nor the difference between light- and mass-weighted estimates are likely to explain this offset, as both have little impact at high masses (\citealp{Trussler2020, Zibetti2026}; \citetalias{Mattolini2025}).
The averaging scheme (mean $[\mathrm{M/H}]$ vs. logarithm of the mean $Z_\star$) may contribute, but cannot be the sole cause, since studies adopting a similar scheme recover the standard metallicity scale \citep[e.g.][]{Gonzalez-Delgado2014}.
A further, possibly dominant, contribution may come from the choice of SPS models: \citetalias{Mattolini2025} show that different assumptions (stellar libraries, isochrones, SFHs, and CEHs) can shift the MZR by up to $\sim0.25~\dex$.

Despite the differences in normalisation, this broad agreement on the existence of a sFMR in the local Universe, supports a scenario in which galaxies mainly evolve smoothly and coherently over long timescales, as physical parameters related to widely different timescales appear to be linked together.

\subsection{Comparing gas-phase and stellar FMRs}
\begin{figure*}[h!]
\centering
\includegraphics[width=\hsize]{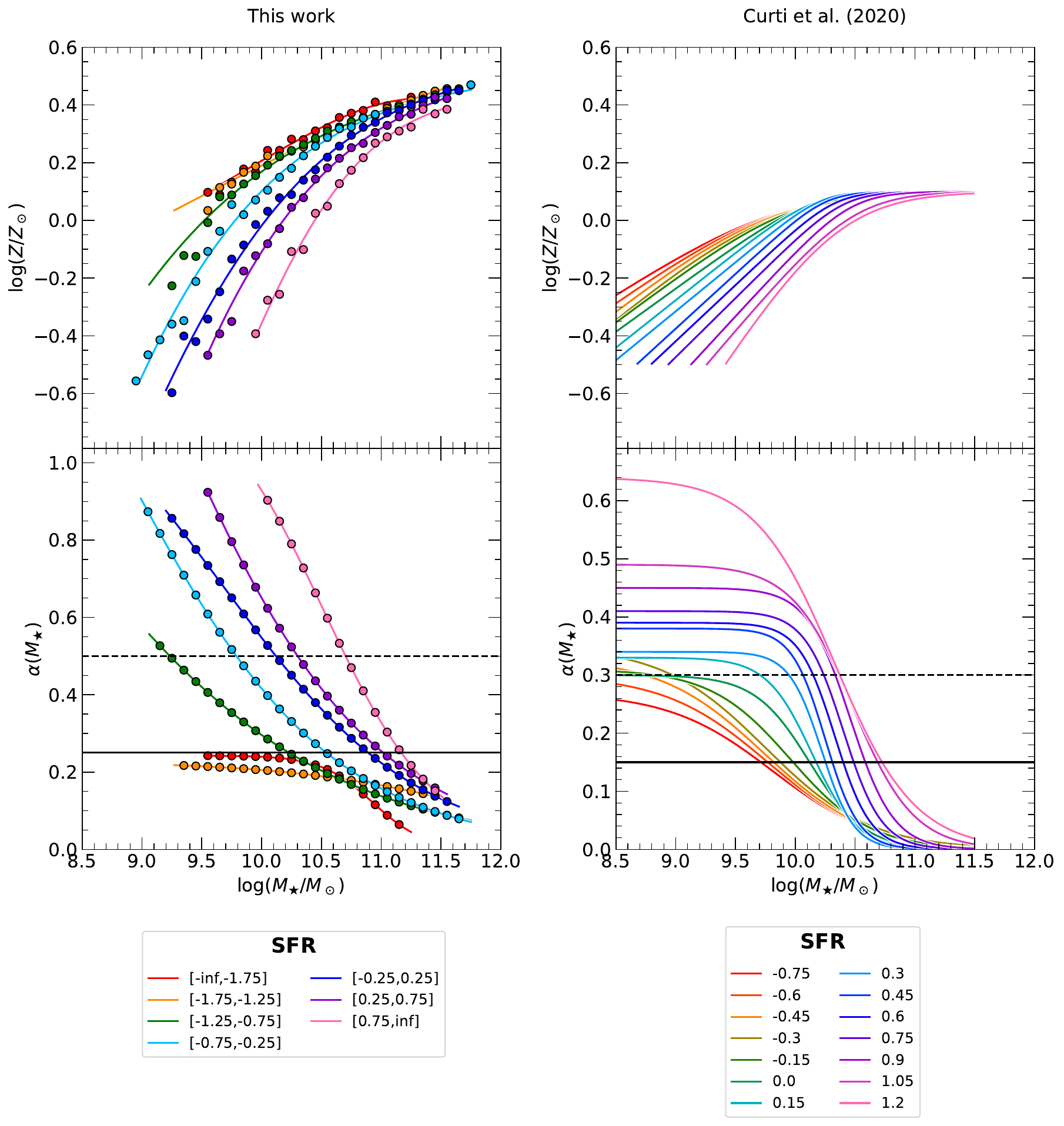}
  \caption{\textit{Top left}: Best fit of the stellar-MZRs from this work (filled circles) using the functional form given in Equation \ref{eq:Curti_rel}, which has been adapted from \citet{Curti2020} (solid lines).
  \textit{Bottom left}: Logarithmic derivative of the fitted stellar-MZRs with respect to stellar mass. 
  Solid lines show the continuous derivative of the fits, filled circles denote the derivative evaluated in $0.1~\dex$-wide bins of stellar mass.
  The horizontal black solid line identifies the threshold $\rm \alpha_{flat}$ used to define the turnover mass $M_{\rm \alpha~flat}$ of the stellar-MZRs.
  The horizontal black dashed line identifies a value twice $\rm \alpha_{flat}$.
  \textit{Top right}: Fit of the gas-MZRs from \cite{Curti2020} (see their Fig. 9).
  \textit{Bottom right}: Logarithmic derivative of the fitted gas-MZRs from \cite{Curti2020} with respect to stellar mass.
  The horizontal black solid line identifies the threshold ($\rm \alpha_{flat}$) used to define the turnover mass $M_{\rm \alpha~flat}$ of the gas-MZRs.
  The horizontal black dashed line identifies a value twice $\rm \alpha_{flat}$.}
    \label{fig:gFMR_sFMR_comparison}
\end{figure*}
\begin{figure*}
\centering
\includegraphics[width=\hsize]{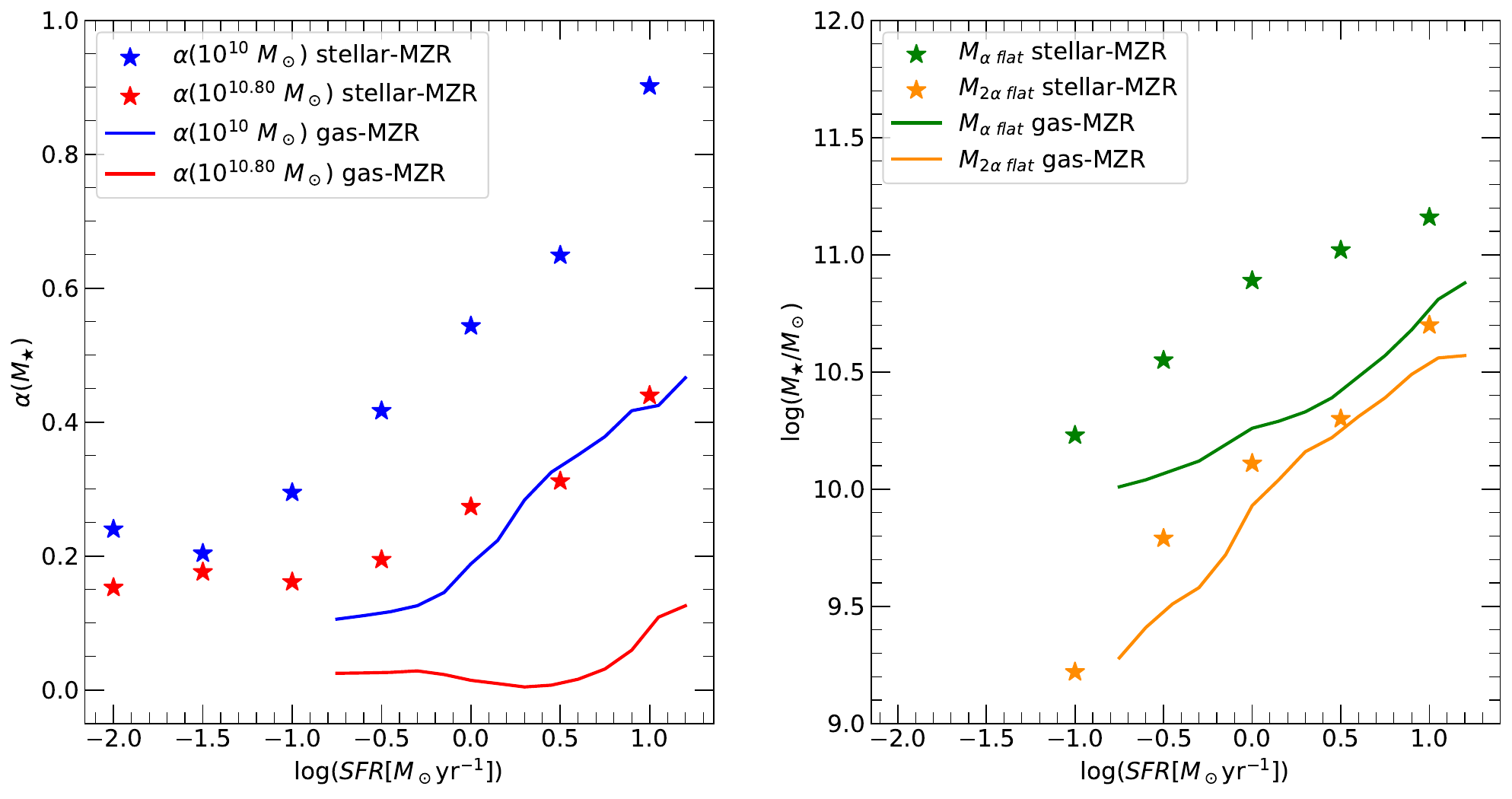}
  \caption{\textit{Left:} slope of the gas- (lines) and stellar-MZRs (stars) as a function of SFR. Blue and red colours identify the slopes associated to galaxy masses of $10^{10}~\Msun$ and $10^{10.80}~\Msun$, respectively.
  \textit{Right:} characteristic masses as a function of SFR estimated from the gas- (lines) and stellar-MZR (stars). 
  Green and orange colours identify the masses at which the slope becomes shallower than $\alpha_{\rm flat}$ and $2\alpha_{\rm flat}$, respectively.
  The figure shows that the gMZRs flatten and saturate to a maximum $Z_{\rm gas}$ at high $M_\star$, whereas the sMZRs show no such saturation ($\rm \alpha_{gas}=0$ and $\rm \alpha_\star\neq0$ in the left panel).
  Additionally, the larger separation between $M_{2,\alpha,\mathrm{flat}}$ and $M_{\alpha,\mathrm{flat}}$ for the sMZRs indicates a slower decrease of their slope with increasing mass compared to the gMZR.
  }\label{fig:Slopes_masses}
\end{figure*}
Our stellar FMR exhibits qualitatively similar trends to the gas-phase FMR \citep[e.g.,][]{Ellison2008, Lara-Lopez2010, Mannucci2010, Salim2014, Cresci2019, Curti2020}. 
While both relations imply that, at a fixed galaxy mass, more actively star-forming galaxies have lower gas- and stellar-metallicities, their shapes differ.
For each bin of star formation activity, the gMZRs are well described by a power-law at low masses, that transitions to a flat relation above a characteristic mass \citep[see eq. \ref{eq:Curti_rel} or eq. 2 in][]{Curti2020}. 
In contrast, our sMZRs do not follow a flat relation at high masses, rather $Z_\star$ keeps increasing with mass.

To provide a quantitative comparison between the gas- and stellar-FMRs, independent of their normalisation, we study the slope of each MZR as a function of mass.
We then compare our stellar MZRs with the most recent analysis of SDSS DR7 galaxies of \cite{Curti2020}, based on strong-line metallicity calibrations \citep[][]{Curti2017}.
In order to compute the slope, defined as the logarithmic derivative $\frac{\mathrm{d}\log(Z_\star)}{\mathrm{d}\log(M_\star)}$, we rely on the analytic fit to the datapoints using the the double power-law parametrisation introduced by \cite{Curti2020}:
\begin{equation}\label{eq:Curti_rel}
    \log(Z_\star)=Z_0-\gamma/\beta\log\left(1+\left(\frac{M}{M_0}\right)^{-\beta}\right)~~~.
\end{equation}
For the sMZRs we perform least-square fits to the median sMZR in the different bins of SFR, while for the gMZRs we adopt the parameters of the fit provided by \cite{Curti2020} (see their tab. 5).
Fig. \ref{fig:gFMR_sFMR_comparison} displays the results of the fit for our stellar-MZRs (top-left panel) alongside the gas-MZRs derived by \cite{Curti2020} (top-right panel).
In each panel the colours are associated to different bins of star formation activity.
In the top-left panel, points represent the stellar-MZR evaluated in $0.1~\dex$-wide stellar mass bins (restricted to bins containing at least $25$ galaxies), while solid lines denote the corresponding best fits.
The bottom panels of Fig. \ref{fig:gFMR_sFMR_comparison} displays the logarithmic derivatives of the gas- (right) and stellar-MZRs (left), representing the slopes ($\alpha$) of the relations as a function of mass.

From the fits shown in the top panels of Fig. \ref{fig:gFMR_sFMR_comparison} we note that, at high masses, the stellar-MZRs are not fully described by the asymptotic flat regime of the \cite{Curti2020} relation, as the stellar metallicity always increase (as shown also by the positive slopes in the bottom-left panel of Fig. \ref{fig:gFMR_sFMR_comparison}).
Consequently, differently to the gas-phase MZRs, this behaviour prevents a reliable estimation of the characteristic turnover mass directly from the parameters of the fit.
In order to provide a consistent analysis of our stellar-MZRs with the gas-MZRs, we empirically define a characteristic flattening mass $M_{\alpha_\mathrm{flat}}$, which corresponds to the mass at which the MZR becomes flatter than a given slope $\alpha_\mathrm{flat}$.
Considering the shapes of the different MZRs and the different dynamical ranges of metallicity for the stars and the gas ($\sim1~\dex$ for $Z_\star$ and $\sim0.6~\dex$ for $Z_{\rm gas}$), the thresholds adopted for the stars and gas are of $\alpha_\mathrm{flat~star}=0.25$ and $\alpha_\mathrm{flat~gas}=0.15$ , respectively\footnote{While the arbitrary choice of the threshold $\alpha_\mathrm{flat}$  affects the determination of the characteristic transition mass, for a sensible range of $\alpha_\mathrm{flat}$ the trends as a function of SFR and the relative differences between gFMR and sFMR are robustly reproduced.}.
The chosen thresholds are indicated by the horizontal black solid lines in the bottom left and right panels of Fig. \ref{fig:gFMR_sFMR_comparison} for the stars and gas, respectively.

Additionally, we compare the slopes at two values of stellar mass for the MZR in each bin of star-formation activity. 
i) The slope at  $M_\star=10^{10}~\Msun$ is used to quantify the slope of the MZRs at the low-mass end.
This is the lowest mass that retains a statistically sufficient population in all $M_\star$-SFR bin (see the left panel of Fig. \ref{fig:FMR_sfr_ssfr}).
ii) The slope at $M_\star=10^{10.80}~\Msun$ quantifies the shape of the MZRs at the transition between the young-galaxy dominated regime and the one dominated by the old galaxies (as derived in \citetalias{Mattolini2025}).

The left panel of Fig. \ref{fig:Slopes_masses} compares the slopes ($\alpha$) for the gas-MZRs (lines) and stellar-MZRs (stars) for the different bins of SFR.
Blue and red lines/stars indicate the slopes at masses of $10^{10}~\Msun$ and $10^{10.80}~\Msun$, respectively.
The right panel of Fig. \ref{fig:Slopes_masses} displays the characteristic masses for  $\alpha_\mathrm{flat}$ and $2\,\alpha_\mathrm{flat}$, from the gas-phase MZRs (lines) and stellar-MZRs (stars).
Green and orange lines/stars indicate the masses at which $\alpha=\alpha_{\rm flat}$ and $\alpha=2\alpha_{\rm flat}$, respectively.
Note that the two least actively star-forming bins do not appear in the right panel of Fig. \ref{fig:Slopes_masses} because the slope of the corresponding MZRs never rise above the threshold $\alpha_{\rm flat}$ (see also Fig. \ref{fig:gFMR_sFMR_comparison}).
From the left panel of Fig. \ref{fig:Slopes_masses} we find that at any given SFR, the gMZRs exhibit shallower slopes than the sMZRs. In particular, at $M_\star=10^{10.80}~\Msun$ the slopes of all gMZRs are below the flatness threshold of $\alpha_{\rm flat~gas}=0.15$, irrespective of the SFR.
A comparison of the slopes at $M_\star=10^{10}~\Msun$ (blue lines/stars) and $M_\star=10^{10.80}~\Msun$ (red lines/stars) reveals that the gMZRs undergo a larger relative change in slope between these two masses, with respect to the sMZRs.
This is evident also from the right panel of Fig. \ref{fig:Slopes_masses}, where the characteristic masses $M_{\alpha\,\mathrm{flat}}$ and $M_{2\,\alpha\,\mathrm{flat}}$~display a larger relative offset for the stellar MZR than for the gas-phase one at all SFRs.
In summary, a direct comparison of the gas- and stellar-MZRs across SFR bins reveals qualitatively similar trends but also relevant differences: the gMZRs flatten at high $M_\star$, saturating to a maximum $Z_{\rm gas}$, whereas the sMZRs show no such saturation -- within each SFR bin, $Z_\star$ continues to rise with $M_\star$, and this rise becomes progressively steeper for more star-forming galaxies.

The different behaviour of the gas- and stellar-FMRs is a consequence of the different physical quantities they trace.
The gas-phase metallicity reflects the present-day chemical composition of the ISM and can rapidly respond to gas inflows, outflows, and metal pollution by supernovae and stellar evolution.
In contrast, the stellar metallicity is the cumulative record of a galaxy's chemical enrichment history, encoding the metallicity of the gas from which subsequent generations of stars are formed.

Focusing on massive galaxies ($M_\star \gtrsim 10^{10.8}\,\mathrm{M_\odot}$), the flattening of the gas-phase MZR at high stellar masses is commonly interpreted within the gas-regulator model framework as the result of an equilibrium between the dilution of the ISM by metal-poor gas inflows, its enrichment through stellar nucleosynthesis, and metal-loaded outflows generated by AGN feedback \citep[e.g.,][]{Dave2011, Dayal2013, Lilly2013, DeRossi2015, Curti2020}.
Over the same range of masses, $Z_{\star}$ still displays a positive trend with $M_\star$. This can be interpreted as an indication that the chemical enrichment history of massive galaxies systematically depends on stellar mass: despite reaching the same endpoint, the most massive galaxies enrich a larger fraction of their stars to a higher degree.
These differences in chemical enrichment history depend not only on the final stellar mass, but also on the current SFR, as apparent from the different sMZRs as a function of star formation activity.
Notably, $M_\star$ and $Z_\star$ are integrals over a galaxy's full evolutionary history, while $Z_{\rm gas}$ and SFR are quasi-instantaneous properties. All in all, this interdependency of physical parameters that probe widely different timescales implies that the chemical enrichment history of massive galaxies is tightly connected to their SFH over a full range of timescales, likely thanks to common regulation mechanisms.

\subsection{Comparison with model predictions}
The presence of a dependence between galaxies' $Z_\star$ and SFR represents a benchmark test for theoretical models of galaxy formation and evolution (e.g. cosmological simulations, semi-analytic models). 
The secondary dependence of $Z_\star$ on SFR, which gives rise to the so-called stellar-FMR, is predicted by semi-analytic models (SAMs).
The sFMR predicted by the GAEA SAM \citep[][]{Hirshmann2016, Fontanot2017, Fontanot2021} shows similar trends compared to our findings, yet with differences in the normalisation, shifting the GAEA relations $\sim0.3~\dex$ towards lower metallicities\footnote{
Note that a proper comparison between observationally inferred light-weighted quantities and theoretical predictions requires full forward modelling \citep[see][]{Nelson2018}. 
This is particularly relevant for the choice of SPS models used to derive stellar population properties, as different models could yield result with differences of $\sim0.25~\dex$ in $Z_\star$ estimates (see also Section 5.1 of \citetalias{Mattolini2025}).
}.
As discussed in \cite{DeLucia2020}, the emergence of a dependence of gas-phase metallicity on galaxies' SFR (which, in turn, could lead to the similar dependence of $Z_\star$ with SFR found in \citealp{Fontanot2021}) critically depends on the implementation of: i) a non-instantaneous gas-recycling \citep[][]{DeLucia2014}, ii) hydrodynamical-based stellar feedback \citep[][]{Hirshmann2016, Fontanot2017}, and iii) a self-consistent partition of the gas in the atomic and molecular components \citep[][]{Xie2017}. 
This further highlights the importance of  robust empirical determinations of these relations to constrain the physical processes that regulate the baryon cycle.

Similarly to SAMs, hydrodynamical cosmological simulations, such as Illustris \citep[][]{Vogelsberger2014, Genel2014}, IllustrisTNG \citep[][]{Weinberger2017, Springel2018, Pillepich2018}, and EAGLE \citep[][]{Schaye2015, Crain2015}, predict the existence of a sFMR, albeit with differing shapes and normalisations \citep[][]{Garcia2024}.
As discussed in \cite{Garcia2024}, the emergence of a $M_\star-\rm SFR-Z_\star$ relation in all three simulations could stem from the treatment of the star-forming ISM with steady (non-bursty) stellar feedback, which in turn results in smooth SFHs, as also suggested by our results. The best agreement of our data is found with Illustris, which, compared to IllustrisTNG, implements weaker stellar feedback at low redshift \citep[see Appendix A of][]{Weinberger2017}.
This weaker feedback results in a reduced suppression of star formation, which, in turn, produces a steeper sMZR, lower stellar metallicities at the low-mass end, and a SFR dependence of the sMZR that more closely resembles our observations \citep[see also figure 3 of][]{Garcia2024}.

Further support to the idea that galaxies' SFHs are shaped predominantly by long-timescale processes (lasting several Gyr), supplemented by minor bursts of star formation on short-timescales, is lent by the small scatter about the SFMS, as shown by \cite{Matthee2019}.  Recent observational results by \cite{Camps-Farina2026} support this scenario, finding that for a sample of $\sim 9\,000$ local galaxies, the scatter around the SFMS is driven by a combination of short-term ($\lesssim100~\Myr$) fluctuations and long-term ($\gtrsim1~\Gyr$) variations in star formation.

\subsection{Redshift evolution of the stellar FMR}
\begin{figure}
\centering
\includegraphics[width=\hsize]{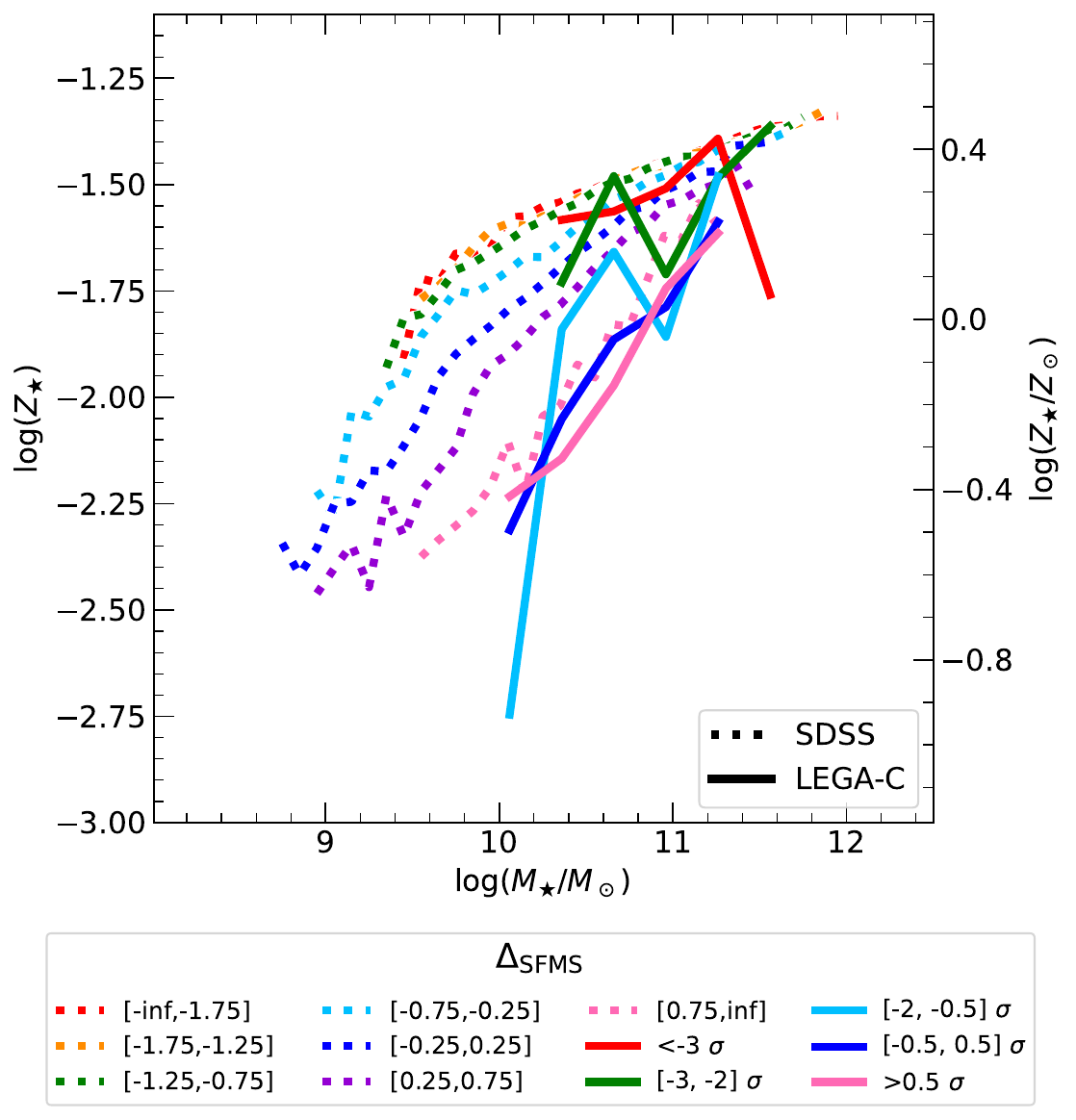}
  \caption{Comparison between the local Universe stellar-FMR (dotted lines) and the $z\sim0.7$ stellar-FMR (solid lines) from \cite{Gallazzi2025PaperII}.
  The differences between local and higher redshift relations increase with increasing star formation activities, with higher redshift relations displaying lower normalisations.}\label{fig:Comparison_Gallazzi2026}
\end{figure}
While both observations and theoretical models indicate that the gas-phase FMR does not evolve significantly up to $z\sim2.5$ \citep[e.g.,][]{Mannucci2010, Curti2020}, predictions from GAEA suggest a significant redshift evolution for the stellar FMR; specifically a decrease in its normalisation at higher redshifts \citep[][]{Fontanot2021}. 
To test for this evolution of the sFMR, in Fig. \ref{fig:Comparison_Gallazzi2026} we compare our local relation (dotted lines) with recent results from \citet{Gallazzi2025PaperII} (solid lines), which are based on a sample of $552$ LEGA-C galaxies at $z\sim0.7$ \citep[][]{vanderWel2016} and utilise the same SPS models and Bayesian inference framework employed in this work.
Note that the bins of distance from the SFMS in \cite{Gallazzi2025PaperII} are defined relative to the scatter $\sigma=0.28$ around the SFMS.
Consequently, there is not a direct correspondence with the bins defined in this work and those defined in \cite{Gallazzi2025PaperII}.

Despite the smaller statistics, and the different bins of distance from the SFMS, the sFMR at $z\sim 0.7$ exhibits trends and shapes similar to the one we observe in the local Universe.
Nonetheless, in accordance with prediction from GAEA, we find differences in the normalisations of the MZRs, in that galaxies at higher redshift have lower stellar metallicities compared to local ones for a given stellar mass and a given distance from the SFMS at the corresponding redshift.
The differences found comparing LEGA-C and SDSS galaxies are of $\sim0.1~\dex$ for the galaxies in the least star-forming bins.
With increasing star formation activity differences increase to $\sim0.25~\dex$.
This comparison indicates that, at fixed stellar mass, the population of more actively star-forming galaxies experience a stronger metallicity evolution across cosmic time \citep[as discussed in sects. 4 and 5 of][]{Gallazzi2025PaperII}.
The similarity of the sFMR in the SDSS and in the LEGA-C samples indicates that the coupling of SFHs with the chemical enrichment history was already in place $\sim6~\Gyr$ ago and has been maintained since then.

\section{Summary and conclusions}\label{sec:Conclusions}
In this paper we have analysed the secondary dependence on star formation activity of the relation between stellar mass and stellar metallicity in the local Universe. We used a sample of 74\,350 SDSS-DR7 galaxies whose stellar population parameters were estimated by \citetalias{Mattolini2025} from high-S/N, aperture-corrected spectral indices and photometry, using state-of-the-art stellar population models. SFRs were taken from the GSWLC catalogue \citep{Salim2016}.

We find that, at fixed stellar mass, the stellar metallicity of local galaxies anticorrelates with SFR, and we quantified this dependence through a partial correlation analysis.
By binning galaxies according to their distance from the star-forming main sequence (SFMS), we showed that the secondary dependence on SFR gives rise to distinct MZRs, which define a surface in the three-dimensional space of $M_\star$, $Z_\star$, and SFR: the stellar fundamental metallicity relation (sFMR).
More actively star-forming galaxies follow steeper MZRs with lower normalisation than less active galaxies.
Consistent results are obtained when binning galaxies by SFR or sSFR instead of distance from the SFMS.
Despite the large uncertainties on individual $Z_\star$ estimates, the large SDSS sample enables a robust determination of the sFMR.
Our analysis confirms the secondary dependence of stellar metallicity on SFR recently reported by \citet{Looser2024} from MaNGA integral-field spectroscopy, using a sample $\sim10$ times larger.
In addition, our analysis includes the recent aperture corrections of \citet{Zibetti2026} and a detailed assessment of the systematics associated with SPS model assumptions \citepalias[see][]{Mattolini2025}.

The gas-phase and stellar FMRs are qualitatively similar in shape. Quantitatively, however, the gas-phase MZRs are systematically flatter than the stellar MZRs in the same SFR bins: the gas-phase MZR flattens at high masses, whereas $Z_\star$ keeps increasing with $M_\star$, albeit with a progressively shallower slope.

Both $Z_\star$, which traces the integrated chemical enrichment history, and the gas-phase metallicity, which traces the present-day enrichment of the ISM, depend on SFR. This suggests that the chemical evolution of galaxies is smoothly coupled with their SFH over timescales of several Gyr. In other words, the current star formation activity of a galaxy is tightly connected to its past SFH and chemical enrichment history, disfavouring a dominant role of stochastic episodes.

Hydrodynamical simulations and semi-analytic models of galaxy formation predict sFMRs qualitatively similar to the observed one, and also predict that the sFMR evolves with redshift. Comparing our local relation with recent results at $z\sim0.7$ \citep[][]{Gallazzi2025PaperII}, obtained with consistent SPS models and inference, we find that such evolution is indeed present.

The existence at $z\sim0.7$ of an sFMR qualitatively similar to the local one suggests that the smooth coupling between SFH and chemical enrichment was already in place at earlier times and has been maintained over the past $\sim6~\Gyr$. Tracing the evolution of these scaling relations back in time is essential to constrain the evolutionary paths of individual galaxies, and our results provide a local benchmark for such studies. Upcoming intermediate-redshift spectroscopic surveys (e.g. WEAVE-StePS, \citealp{Iovino2023a}; 4MOST-StePS, \citealp{Iovino2023b}) will bridge the gap between $z\sim0$ and $z\sim0.7$ with more continuous redshift coverage, while deep near-infrared (NIR) surveys \citep[e.g. MOONRISE;][]{Cirasuolo2020} will extend this analysis towards cosmic noon.

\section*{Data availability}
The table containing the observed and estimated properties of the duplicate-combined SDSS DR7 galaxies data used in this work are publicly available in electronic form at the CDS anonymous ftp to cdsarc.u-strasbg.fr (130.79.128.5), via http://cdsweb.u-strasbg.fr/cgi-bin/qcat?J/A+A/, and on the BaStA website at \url{https://www.basta.inaf.it/}.

\begin{acknowledgements} 
We thank the anonymous referee for his/her constructive feedback, which helped improve the presentation and discussion of our results.
This paper and related research have been conducted during and with the support of the Italian national inter-university PhD programme in Space Science and Technology.
SZ acknowledges support from the INAF-Minigrant-2023 “Enabling the study of galaxy evolution through unresolved stellar population analysis” 1.05.23.04.01.
LSD acknowledges receiving funding from the European Union Horizon Europe Research and Innovation Action under grant agreement no. 101183153-WST. Views and opinions expressed are however those of the author(s) only and do not necessarily reflect those of the European Union or the European Research Executive Agency (REA). Neither the European Union nor the REA can be held responsible for them.
\end{acknowledgements}

\bibliographystyle{aa}
\bibliography{Bibliografia}

\begin{appendix}
\section{Uncertainties on stellar-metallicities and the impact of signal-to-noise selection}\label{app:MZRs_bins_SFR}
\begin{figure*}[h!]
\centering
\includegraphics[width=\hsize]{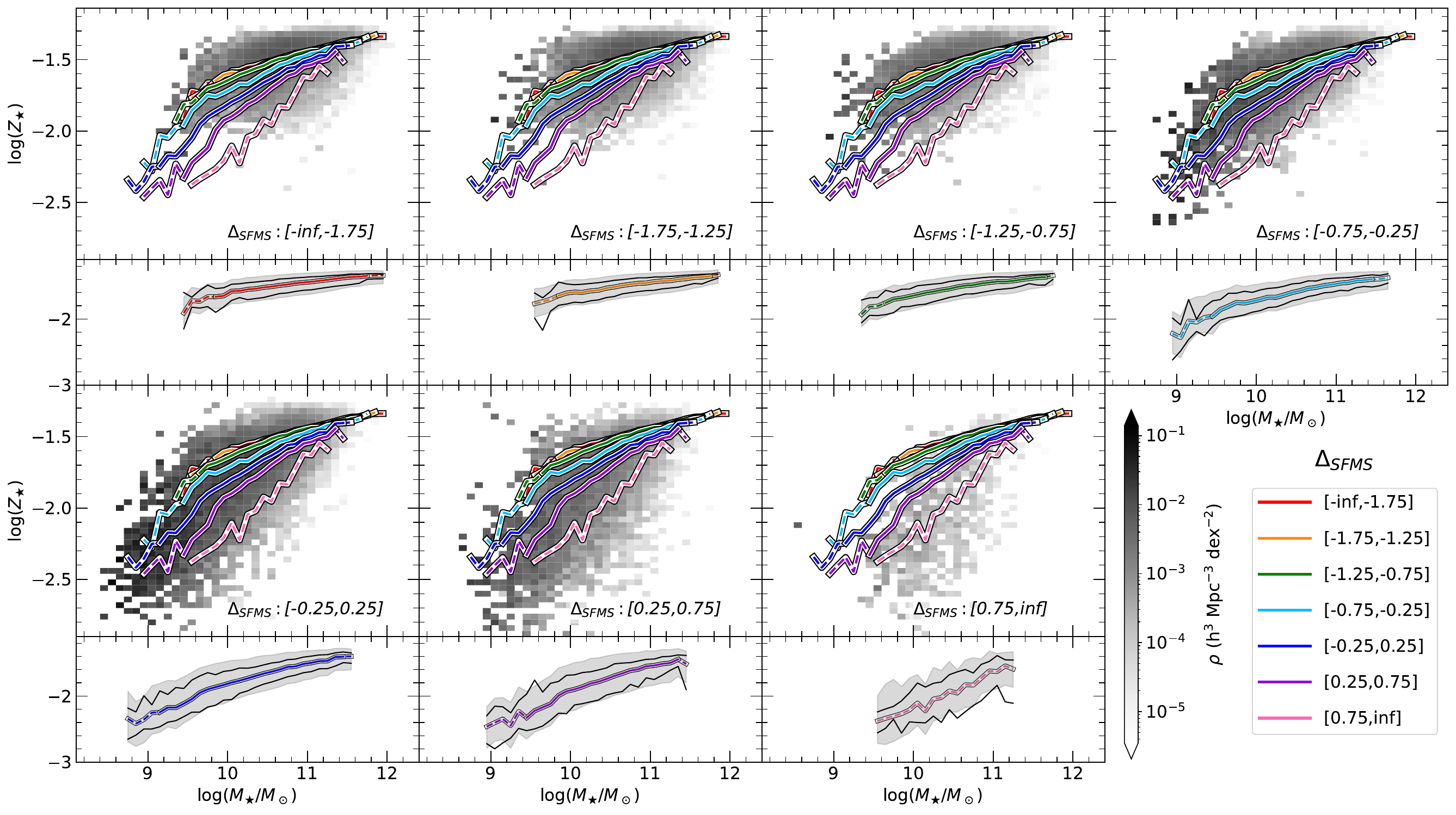}
  \caption{The coloured lines identify the median MZRs of galaxies in each bin of $\DSFMS$. Solid and dashed lines identify those bins with at least 50 and 10 galaxies per bin, respectively.
  The grey 2D histogram represents the number density of the complete galaxy sample in the $M_\star-Z_\star$ plane.
  In each sub-panel, the shaded area identifies the uncertainty associated with the metallicity estimates, and the black lines identify the $16^{\rm th}$ and $84^{\rm th}$ percentiles of the corresponding mass-metallicity distribution.}
  \label{fig:FMRRs_sfr_bins}
\end{figure*}
As presented in Sect. \ref{sec:Results}, within each bin of $\DSFMS$, the uncertainties on the $Z_\star$ estimates of individual galaxies are generally larger than the spread of the corresponding MZR. This indicates that part of the error budget of $Z_\star$ is systematic and does not broaden the MZR. Additionally, the uncertainty on the median of the MZR is significantly smaller than the spread of the relation. In this appendix, we detail the estimation of the uncertainties on the median MZRs and their spread. We also discuss possible physical motivations for the trends of individual metallicity uncertainties and the spread of the MZRs with stellar mass and star formation activity. Finally, we analyse the MZRs in bins of $\DSFMS$ obtained with a stricter $\rm S/N$ selection to test the robustness of our results.

Figure \ref{fig:FMRRs_sfr_bins} displays the MZRs in each bin of $\DSFMS$, from the least actively star-forming (top left panel) to the most actively star-forming one (bottom right panel).
In the upper part of each sub-panel we show the median relations with the coloured lines (using the same colour scheme of Fig. \ref{fig:FMR_dist_sfr}), overplotted to the 2D histogram of the number density distribution of galaxies.
In the lower part, the median uncertainties of $Z_\star$ estimates of individual galaxies in each mass-bin are shown with the shaded area.
The spreads of the mass-metallicity distribution are identified with the black lines.
From Fig. \ref{fig:FMRRs_sfr_bins} it is noticeable that uncertainties in metallicity estimates depend both on $M_\star$ and $\DSFMS$.
In the lowest mass bin, the median uncertainties on metallicity estimates for the most and the least star-forming galaxies are of $\sim0.35~\dex$ and $\sim0.20~\dex$, respectively.
With increasing $M_\star$, these uncertainties decrease, reaching values of $\sim0.20~\dex$ and $\sim0.15~\dex$, respectively.

The mass dependence of metallicity uncertainties may originate from weaker metal absorption in the observed spectra.
This can be due to two reasons: i) for each bin of $\DSFMS$, due to the MZR, lower-mass galaxies are characterised by lower metallicities, naturally leading to weaker absorption; ii) given the presence of a mass-age relation \citep[e.g.][\citetalias{Mattolini2025}]{Gallazzi2005, Gallazzi2021}, the stellar populations hosted by lower-mass galaxies are also younger, thus characterised by weaker metal absorption when compared to old stellar populations with the same $Z_\star$.
Additionally, at younger ages the degeneracy in spectral properties with respect to SFH and chemical enrichment is more severe \citep[see, e.g., Fig. 40 in][]{Rossi2025}, producing wider posterior PDFs and larger uncertainties.

Given the existence of a positive correlation of both $Z_\star$ and the SFR with $M_\star$, the same reasoning can be used to explain the dependence of the uncertainties of stellar metallicity on galaxies' star formation activity.
Similar dependencies on $M_\star$ and SFR are found when analysing the spread of the MZRs.

We also calculated the uncertainties on the median estimates of the MZRs, defined as $\sigma_Z=\Delta/\sqrt{N}$, where $\Delta$ is half the range between the $\rm 16^{th}$ and $\rm 84^{th}$ percentiles of the metallicity distribution, and $N$ is the number of galaxies in the selected bin.
We found that $\sigma_Z$ shows similar trends with $M_\star$ and star formation activity, with uncertainties decreasing as $M_\star$ increases or $\DSFMS$ decreases. 
These uncertainties in the median MZRs are mostly below $\sim 0.01~\rm dex$, reaching the largest values of $\sim 0.04~\rm dex$ only in the most active star-forming bin, which also corresponds to the bin with the poorest statistics.

\begin{figure}
\centering
\includegraphics[width=\hsize]{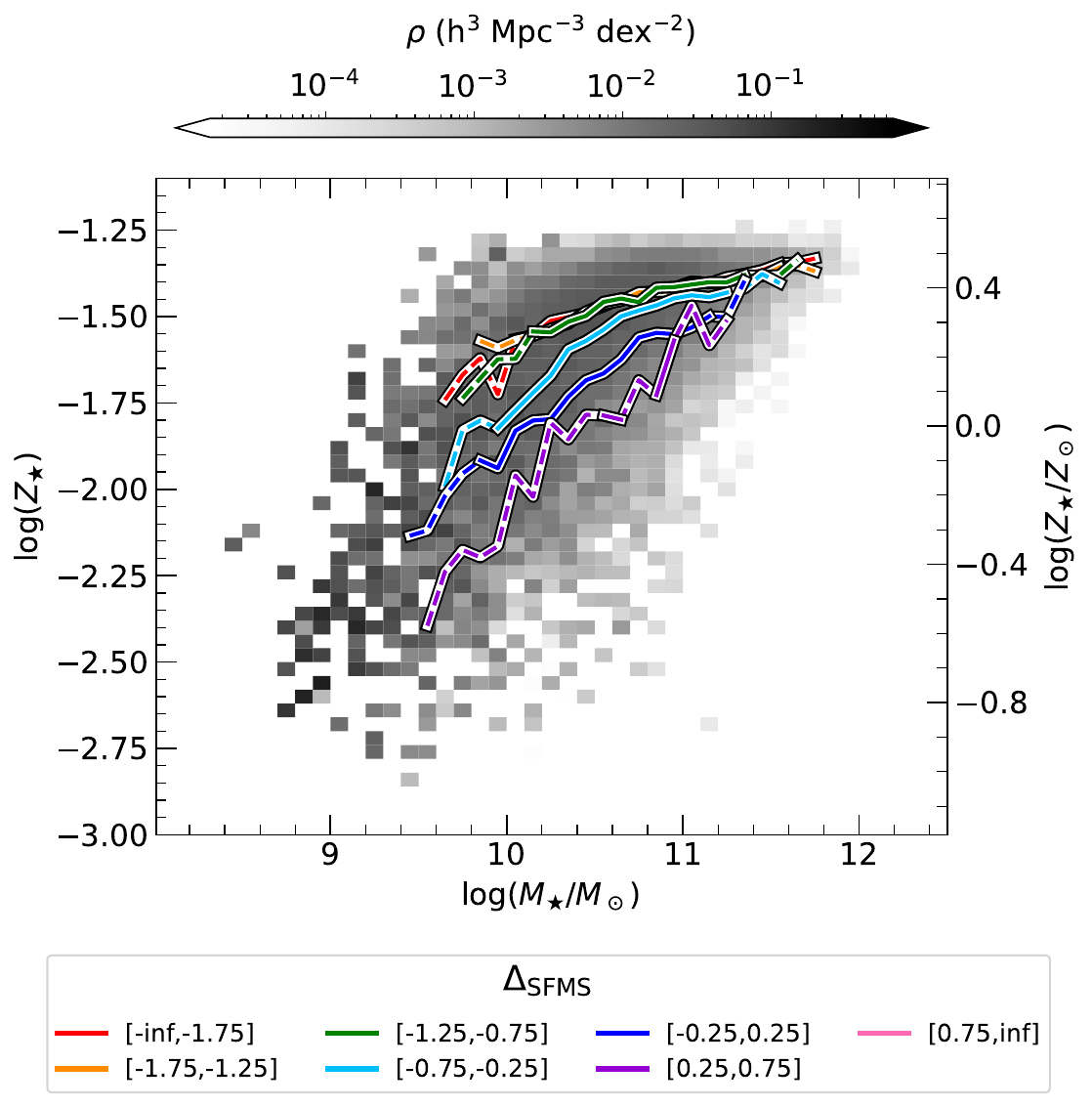}
  \caption{The local Universe stellar-FMR for galaxies with $\rm S/N\geq30$.
  The coloured lines identify the median MZRs of galaxies in different bins of SFR. The solid (dashed) lines identify those bins with at least 50 (10) galaxies per bin.
  The grey 2D histogram represents the number density of the complete galaxy sample with $\rm S/N\geq30$ in the $M_\star-Z_\star$ plane.}
  \label{fig:FMR_SNR30}
\end{figure}

We tested the robustness of our results by analysing the sFMR for a sample of galaxies with $\rm S/N\geq30$.
In Fig. \ref{fig:FMR_SNR30}, the coloured lines identify the MZRs for different bins of $\DSFMS$ (following the colour scheme of Fig. \ref{fig:FMR_dist_sfr}).
The number density distribution of galaxies in the $M_\star-Z_\star$ plane obtained by selecting galaxies with $\rm S/N\geq30$ is represented by the underlying grey 2D histogram\footnote{
The statistical corrections have been recomputed consistently with the different S/N cut, following \citetalias{Mattolini2025}.
}.
Restricting the galaxy sample to higher $\rm S/N$, we do not have enough statistics to trace the MZR of the most actively star-forming galaxies and, for masses $M_\star\lesssim10^{10}~\Msun$, we suffer from significant incompleteness.
Nevertheless, the derived sFMR remains generally consistent with the results obtained from the $\rm S/N\geq20$ sample.
The only notable difference is that the higher $\rm S/N$ selection results in a slightly (a few $0.01$ dex) larger separation between the MZRs in the $10^{10}-10^{11}~\Msun$ mass range compared to the lower S/N counterpart. This mainly follows from a small shift of the galaxies with negative $\DSFMS$ towards higher metallicity. This effect could be partly explained by a selection bias, whereby high S/N galaxies tend to have higher mass concentrations and, as a consequence, higher $Z_\star$ \citep[see, e.g.,][]{Zibetti2022}.

\section{Impact of different SFR estimates on the sFMR}\label{app:sfr_comparison}
In this appendix we compare the fiducial results presented in the main text, based on the SFRs of \citet{Salim2016} (\SFRS{} hereafter), with those obtained using the SFRs of SDSS DR8 galaxies derived following \citet{Brinchmann2004} and \citet{Salim2007} (\SFRDR{} hereafter)\footnote{Publicly available at \url{https://www.sdss4.org/dr17/spectro/galaxy_mpajhu/}.}.
We briefly describe how these estimates are derived and then analyse the differences in the resulting trends. We refer the reader to \citet{Brinchmann2004} and \citet{Salim2016} for a detailed comparison of the two SFR estimates.

\SFRDR{} values are derived from emission lines, which provide robust SFRs for star-forming galaxies (classified through the BPT diagram) with sufficiently strong H$\alpha$ emission. For galaxies hosting an AGN or with weak emission lines, \SFRDR{} is instead derived from photometry.
Because of the finite size of the SDSS fibres, H$\alpha$-based SFRs are subject to aperture bias, as they only sample the central regions of galaxies. Following \citet{Brinchmann2004}, aperture corrections are derived from a relation between the in-fibre SFR and broad-band colours, which is then applied to the light outside the fibre to estimate the out-of-fibre SFR \citep[see also][]{Salim2007}.

\begin{figure}
\centering
\includegraphics[width=\hsize]{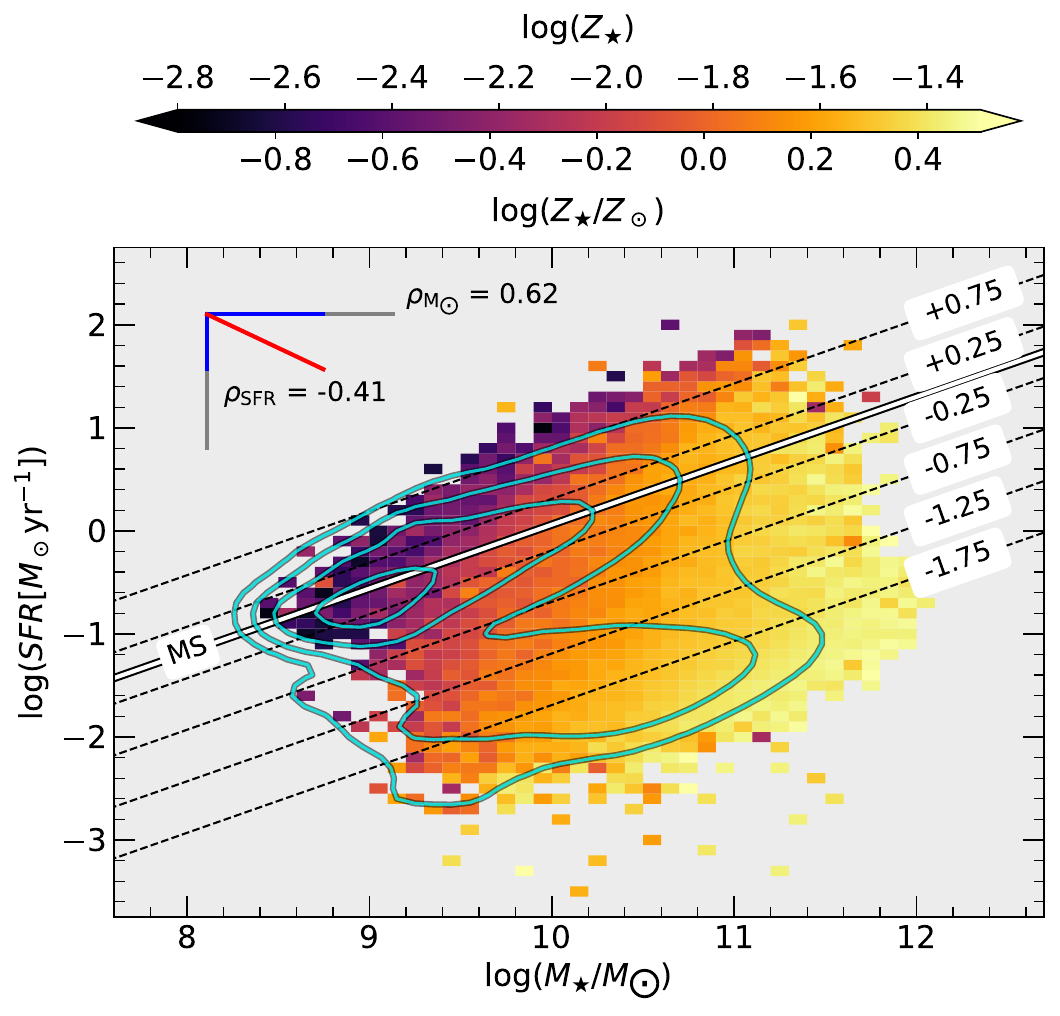}
  \caption{Same as Fig.~\ref{fig:Mstar-SFR_cc_Zstar}, but using \SFRDR{} and the SFMS of Eq.~\eqref{eq:SFMS_fit_DR8}.}
  \label{fig:Mstar-SFR_cc_Zstar_B04}
\end{figure}

Figure~\ref{fig:Mstar-SFR_cc_Zstar_B04} shows the distribution of our sample in the $M_\star$--\SFRDR{} plane, colour-coded by the median $\log Z_\star$ in each bin. As in Fig.~\ref{fig:Mstar-SFR_cc_Zstar}, cyan contours show isodensity levels, the solid white line is our SFMS fit, and the dashed black lines delimit the $0.5~\dex$-wide bins in $\DSFMS$.
For star-forming galaxies around the SFMS, the two SFR estimates yield very similar distributions, consistent with the agreement between the two estimates found by \citet{Salim2016} for this population. Significant differences arise for galaxies with low star formation activity: \SFRDR{} produces a well-defined passive sequence extending from the high-mass end down to $\sim10^9~\Msun$, which is not seen with \SFRS{} (Fig.~\ref{fig:Mstar-SFR_cc_Zstar_B04}). This reflects the large discrepancies between the two estimates for galaxies hosting AGN or with weak emission lines, which lie in and below the green valley \citep{Salim2016}. As a consequence, the fraction of galaxies in the least active $\DSFMS$ bin is $\sim11\%$ with \SFRDR{}, compared with $\sim6\%$ with \SFRS{}.
Since the two estimates agree for star-forming galaxies, the SFMS derived with \SFRDR{},
\begin{equation}\label{eq:SFMS_fit_DR8}
    \log \mathrm{SFR}_\mathrm{MS}(M_\star) = 0.62\,\log(M_\star/\Msun) - 6.14\,,
\end{equation}
is nearly identical to the fiducial one (Eq.~\ref{eq:SFMS_fit}).

\begin{figure}
\centering
\includegraphics[width=\hsize]{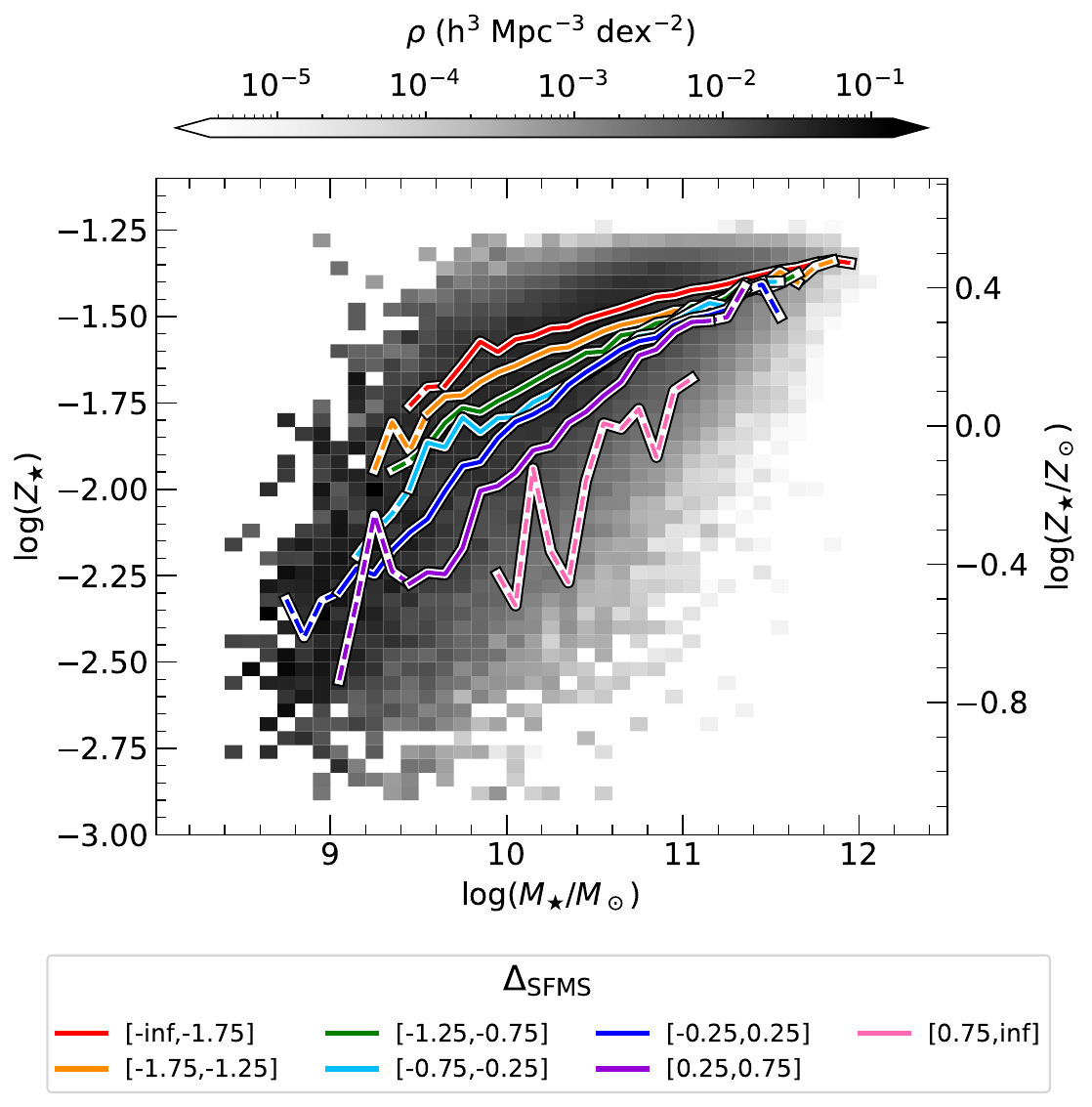}
  \caption{Same as Fig.~\ref{fig:FMR_dist_sfr}, but with galaxies binned by distance from the SFMS of Eq.~\eqref{eq:SFMS_fit_DR8}, using \SFRDR{}.}
  \label{fig:FMR_sfr_DR8}
\end{figure}

The colour map in Fig.~\ref{fig:Mstar-SFR_cc_Zstar_B04} also shows that $Z_\star$ is distributed differently across the $M_\star$--SFR plane: the gradient of $Z_\star$ with SFR is steeper than in Fig.~\ref{fig:Mstar-SFR_cc_Zstar}, particularly for passive low-mass galaxies. Consistently, the partial correlation coefficients, $\rho_{M_\star}=0.62$ and $\rho_{\SFR}=-0.41$, indicate a stronger dependence of $Z_\star$ on \SFRDR{} than on \SFRS{} ($0.52$ and $-0.27$, respectively, see Sect.~\ref{sec:Results}).
Figure~\ref{fig:FMR_sfr_DR8} shows the median MZRs in bins of $\DSFMS$ derived with \SFRDR{}, with the same layout as Fig.~\ref{fig:FMR_dist_sfr}. The resulting sFMR shows the same qualitative trends as the fiducial one. For star-forming galaxies ($\DSFMS\gtrsim-0.25$), the MZRs in corresponding bins are nearly identical in shape and normalisation. Differences arise for galaxies more than $\sim0.3~\dex$ below the SFMS: with \SFRDR{}, the MZR of the most passive bin is $\sim0.05~\dex$ higher than the fiducial one, while in the $[-1.25,-0.75]$ and $[-0.75,-0.25]$ bins it is $\sim0.05~\dex$ lower. As a result, the MZRs of galaxies below the green valley are more widely separated when \SFRDR{} is used.

This larger separation most likely arises from methodological biases. For passive galaxies, \SFRDR{} relies on optical photometry alone, which, owing to the age--metallicity degeneracy, is sensitive to both stellar age, and hence star formation activity, and stellar metallicity. This can introduce a spurious correlation between \SFRDR{} and $Z_\star$. In contrast, \SFRS{} benefits from GALEX UV constraints, which are more directly sensitive to recent star formation. In addition, H$\alpha$-based and photometric SFRs probe different timescales ($\sim10~\Myr$ vs. up to $\sim1~\Gyr$), so that \SFRDR{} is not homogeneous across the star-forming and passive populations, whereas \SFRS{} traces a consistent $\sim100~\Myr$ timescale.
\end{appendix}
\end{document}